# Analysis of Conducted and Radiated Emission on a Self-oscillating Capacitive Touch Sensing Circuit

Subramaniam Saravana Sankar
*Department of Security Engineering*
*Faculty of Applied Informatics,*
*Tomas Bata University in Zlin*
Zlin, Czech Republic
saravana_sankar@utb.cz

Stanislav Kovar
*Department of Security Engineering*
*Faculty of Applied Informatics,*
*Tomas Bata University in Zlin*
Zlin, Czech Republic
skovar@utb.cz

Martin Pospisilik
*Department of Electronics and Measurements*
*Faculty of Applied Informatics,*
*Tomas Bata University in Zlin*
Zlin, Czech Republic
pospisilik@utb.cz

Michael Galda
*Systems Engineering*
*NXP Semiconductors*
Roznov p.R., Czech Republic
michael.galda@nxp.com

*Abstract*—With the advent of smartphones, there has been a recent increase in the use of capacitive touch sensing for various Human Machine Interfaces (HMI). Capacitive-based touch sensing provides higher flexibility and cost-effectiveness than, methodologies such as resistive-based touch sensing. However, Capacitive-based touch sensing is more prone to disturbances such as Electromagnetic interference (EMI) and noise due to temperature variation. This effect becomes more dominating as the sensing excitation frequency increases. Traditional capacitance to digital circuits, such as sigma-delta capacitive sensing, requires multiple clock cycles to measure sensing capacitance, thus necessitating higher frequency operation. In turn, this produces challenges in Electromagnetic Emission while also increasing its susceptibility to EMI, such as false or ghost touch due to exposure of the sensing electrodes to various frequency electric fields. This paper discusses the conducted electromagnetic emission behavior of an external excitation-frequency independent self-oscillating capacitance-to-time converter, where sensing is done with a single clock cycle, and discusses radiated Electromagnetic emission of the touch sensing electrode. The proposed approach is suitable for touch-sensing applications, mainly when used in a noisy EMI environment, such as inside a vehicle within the Automotive industry.

## I. INTRODUCTION

Touch sensing, in general, has been widely used in many applications, particularly in consumer-grade electronics such as laptops, mobile phones, and tablets [1], [2]. Broadly there are two different kinds of touch sensing currently in existence: resistive touch sensing, which works by sensing voltage drop across a resistance caused by the touching action, and capacitive touch sensing, which works by measuring the change in capacitance caused by touch action. Due to its flexibility and cost-effectiveness, capacitive-based touch sensing is widely used in consumer electronics. Capacitive touch sensing is further divided into two types: self-capacitance-based touch sensing and mutual capacitance-based touch sensing, as shown in Fig. 1.

However, capacitive-based touch sensing is more susceptible to both conductive and wireless EMI; this phenomenon becomes more visible when the sensing frequency increases. As an example, the malfunction of a consumer device's touch screen due to small common-mode (CM) currents (few µA) flowing from a power adapter to the Equipment Under Test (EuT) was concluded by the researchers in [3].

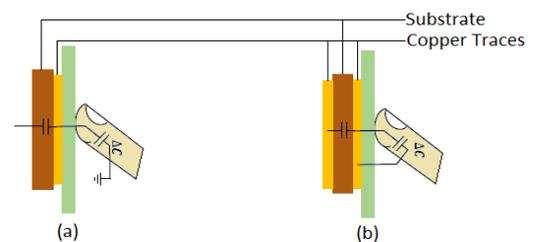

Fig. 1 (a) Self-capacitance-based touch sensor, and (b) mutual capacitance-based touch sensor

Typically, the change in capacitance of capacitive touch sensing is measured by pumping small discreet charge packets to the sensor capacitor and counting the total number against a known capacitor; one such example is the sigma-delta sensing circuit. Therefore, speed and resolution are directly proportional to the operating frequency, making this a drawback concerning EMI [4]–[8].

This paper's research is based on a self-oscillating capacitive sensing circuit where sensing is done with one oscillating cycle, in which the speed and resolution depend on the change in the duty cycle. The presented research analyses the influence of the sensing circuit's oscillating or operating frequency on conducted CM emission over multiple frequencies. To illustrate the benefits of operating the sensing circuit at a lower frequency, this research also compares the near electric field (E-field) emission at the touch sensing electrode excited by driving pulse signal operating at a higher frequency relative to the self-oscillating circuit's operating frequency.

The rest of this paper is arranged as follows: Section II details the working of the sensing circuit test setup within the simulation environment. Section III discusses the analysis of the simulation and experimental procedures. Section IV discusses conclusions on the results.

## II. METHODS AND CONDITIONS FOR CONDUCTED EMI MEASUREMENT ON CAPACITIVE SENSING CIRCUIT

### A. Operation Principle of an Oscillating Capacitive Sensing

An oscillating capacitive sensing circuit is proposed, as shown in Fig. 2. The core of the sensing circuit consists of a relaxation oscillator, similar to the work done in [9]. The following equations give the relationship between the sensing capacitance $C_{sen}$ and the output voltage $V_{do}$ regarding duty cycle and frequency, where $T_{on}$, $T_{off}$, $f_{osc}$ express the on-time, off-time, and the frequency of the oscillation, respectively.

$$K = R_1/R_2 \quad (1)$$
$$J = (C_{sen} - C_{off})/C_{Int} \quad (2)$$
$$T_{on} = R_{ref} * C_{Int} * K \quad (3)$$
$$T_{off} = R_{ref} * C_{Int} * (K - 2J) \quad (4)$$
$$f_{osc} = 1/(T_{on} + T_{off}) \quad (5)$$

From (3), and (4), it is evident that the change in duty cycle to the change in capacitance $C_{sen}$ is inversely proportional to the operating frequency $f_{osc}$, thus requiring higher frequency digital reading at $V_{do}$ to minimize data loss while maintaining reasonable measurement time.

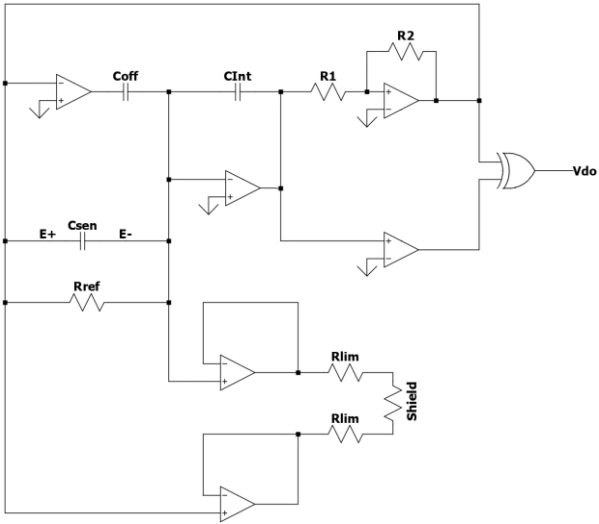

Fig. 2 Schematic of the proposed capacitive touch sensing circuit

However, given the possibility of isolation such as optical coupling at the output with less effort, thus this study does not consider the influence of potential EMI noises due to high-speed data acquisition from the proposed circuit.

### B. EM Simulation parameters and conditions

The proposed sensing circuit is designed within the Electronic Design Automation (EDA) software suite and the same is simulated in quasi-3D EM computing environment for the conducted EMI. The software tool used is CST PCB Studio by Dassault Systèmes Simulia. The problem-solving methodology used is Partial Element Equivalent Circuit (PEEC), in which a 3D multi-layer Printed Circuit Board (PCB) is divided into a mesh of short conductive segments with dielectric areas. Inductors do the magnetic coupling between the coppers and the electric coupling between the conductive copper regions is done by capacitors, which considers the impact of the dielectric areas.

Fig. 3 illustrates the relationship between the applied methodology regarding the quasi-3D simulation and the 2D (Circuit Level) co-simulation within the software. The Line Impedance Stabilization Network (LISN) as shown in Fig. 4, used for measuring the conducted EMI, is modelled similarly to the work done in [10]. TABLE 1 shows some of the critical parameters in this work regarding modelling of the proposed circuit.

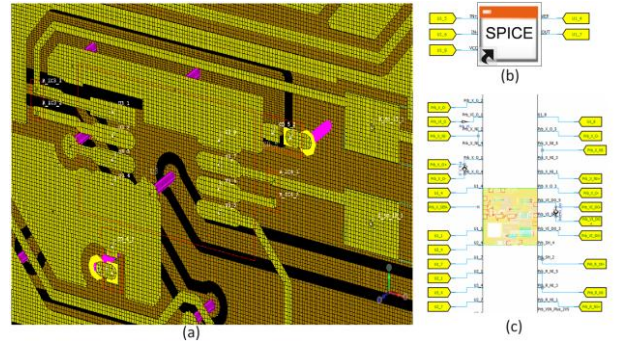

Fig. 3 (a) Shows the mesh view of the PCB. (b) and (c) represents the linkage between the 2D and 3D simulation.

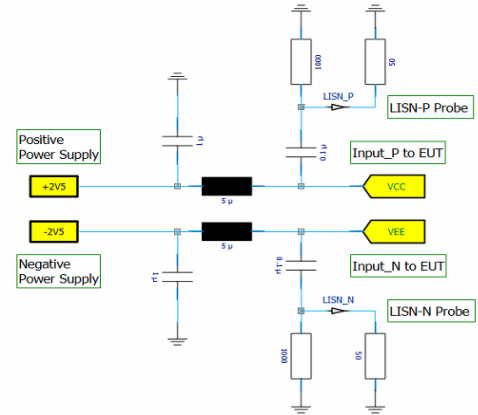

Fig. 4 Simulated LISN network.

TABLE 1. SIMULATION PARAMETERS

| Parameters | Values |
|---|---|
| Total Nets (Unique Electrical Connections) | 40 |
| Power Management | ±2.5V, GND (Reference) |
| Element Size[a] | 0.5mm |
| Capacitive Segments (Signal) | 7704 |
| Capacitive Segments (Reference) | 15771 |
| Inductive Segments (Signal) | 11672 |
| Inductive Segments (Reference) | 30106 |
| Vias Inductive Segments (RL only model) | 69 |

[a]. For faster calculation, element size is adjusted based on $f_{osc}$.

## III. RESULTS AND ANALYSIS

### A. Conducted Emission Simulation Results

The conducted EMI measurement within the simulation is done by measuring the voltage across 50Ω impedance presented to the virtual EMI receiver by the LISN for each supply rail, which can be defined by the (6) and (7), where $V_P$, $V_N$ corresponds to conducted noises at the +2.5V and -2.5V rails respectively; $I_{CM}$, $I_{DM}$ represents common and differential mode currents.

$$V_P = 50 * (I_{CM} + I_{DM}) \quad (6)$$
$$V_N = 50 * (I_{CM} - I_{DM}) \quad (7)$$

The virtual EMI receiver within the simulation is set up with the following provisions [11]:

- the frequency range between 150kHz to 108MHz is selected,
- resolution Bandwidth (RBW) of 9kHz is selected for the entire frequency range,
- Gaussian window function is applied for the RBW, with 1µS width and 95% overlapping.

Fig. 5 and Fig. 6 shows the simulated conducted emission results for touch sensing circuit operating at $f_{osc}$ = 25kHz and $f_{osc}$ = 1kHz respectively.

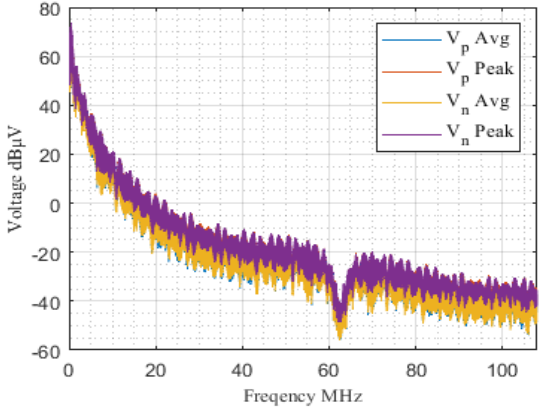

Fig. 5 Simulated conducted emission at $f_{osc}$ = 25kHz

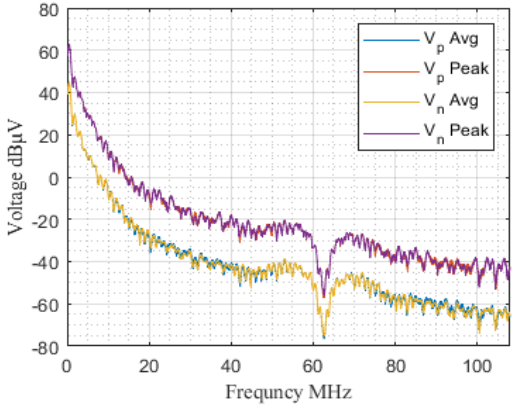

Fig. 6 Simulated conducted emission at $f_{osc}$ = 1kHz

*B. Radiated Emission Experimental Results*

The intensity and frequency of the driving pulse signal from the capacitive sensing circuit predominantly influence the radiated emission from the capacitive touch sensing electrode. The current through the capacitor is governed by the (8), where higher frequency and intensity are preferred to improve sensitivity, enhancing the signal-to-noise (SNR) ratio. However, such methodologies can exacerbate radiated EMI generation due to intense current flow and increase of emission interval due to increased $dV/dt$ and total number of sensing cycles.

$$i(t) = C * \frac{dV}{dt} \quad (8)$$

Fig. 7 shows the experimental arrangements for measuring the near-field emission from the capacitive sensing electrode, similar to the work done by [12]. This study primarily examines how changes in the frequency of the driving pulse to the capacitive touch electrodes, while maintaining a constant peak-to-peak voltage of 5V, impact the near-field emission.

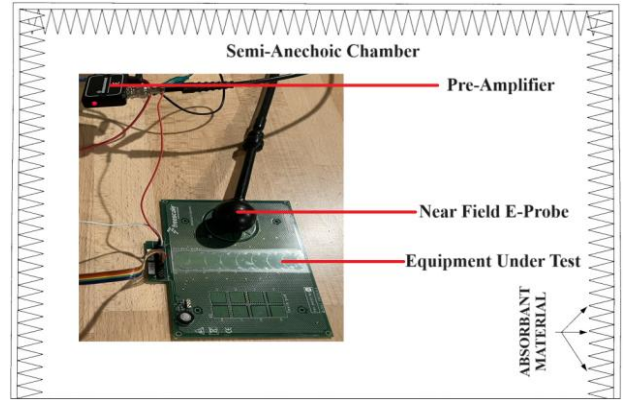

Fig. 7 Measurement configuration for near field emissions

In this test near-field E-probe made by Rohde & Schwarz (Model: HZ-11) with following specifications is used,

- field rejection is 30dB with having 1st resonant frequency >1GHz,
- preamplifier with nominal gain of 38dB is used.

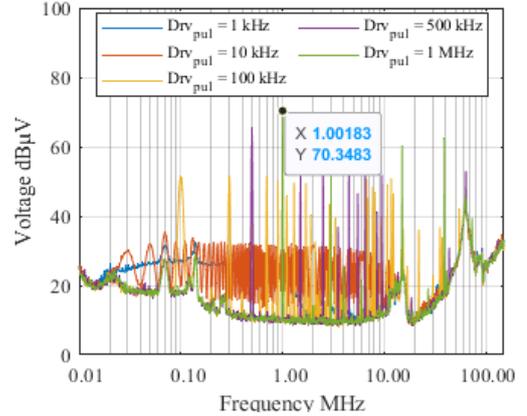

Fig. 8 Measured levels of electric field emissions.

Fig. 8 represents the results from the experiment, where input driving pulse frequency is varied from 1 kHz to 1 MHz, and the results indicate that, as the frequency of operation increases, the associated noise also increases. The results in Fig. 8, also show that the peaks of noise and its harmonics are directly related to the driving pulse frequency of the sensing electrodes, one such corresponding data point for 1 MHz frequency is marked in the graph. This phenomenon can be attributed to the touch electrodes having a poor return path, and thus generates EMI which is similar to the generation of CM EM field due to the capacitive load [13].

IV. CONCLUSIONS

This paper evaluates the performance of a capacitive sensing circuit in terms of conducted EM emission within the simulated environment and investigates the near electric field emission of the capacitive touch sensing electrode, which is driven by various frequency driving pulse signals. The results of the proposed capacitive sensing circuit and its conducted emission simulation indicate that as the frequency at which the charging and discharging of the capacitor, i.e., the frequency of oscillation, increases, the noise associated with the same also increases.

The near-field emission experiment on the sensing electrode shows that the radiated near field EM generation is directly proportional to the frequency of the driving pulses. In many touch-based devices, capacitive touch electrodes occupy a significant physical space closer to sensitive electronics. From the results of this paper, it can be exclaimed that increasing operating frequency of sensing for better performance may exacerbate EMI to the nearby noise-sensitive electronics.

Future research aims to expand the current research, especially to analyze the present properties of capacitive touch sensors and reveal weak parts concerning resistance, radiation, temperature effects, and likewise.

## Acknowledgment

This project has received funding from the European Union's Horizon Europe research and innovation program under the Marie Sklodowska-Curie grant agreement No. 101072881 and UKRI.

## References

[1] Y. Sugita, K. Kida, and S. Yamagishi, "In-Cell Projected Capacitive Touch Panel Technology," IEICE Trans. Electron., vol. E96.C, no. 11, pp. 1384–1390, 2013, doi: 10.1587/transele.E96.C.1384.

[2] T. Wang and T. Blankenship, "Projected-Capacitive Touch Systems from the Controller Point of View," Inf. Disp., vol. 27, no. 3, pp. 8–11, 2011, doi: 10.1002/j.2637-496X.2011.tb00363.x.

[3] Y. Li, S. Wang, H. Sheng, C. P. Chng, and S. Lakshmikanthan, "Investigating CM Voltage and Its Measurement for AC/DC Power Adapters to Meet Touchscreen Immunity Requirement," IEEE Trans. Electromagn. Compat., vol. 60, no. 4, pp. 1102–1110, Aug. 2018, doi: 10.1109/TEMC.2018.2794318.

[4] O. Kanoun, A. Y. Kallel, and A. Fendri, "Measurement Methods for Capacitances in the Range of 1 pF–1 nF: A review," Measurement, vol. 195, p. 111067, May 2022, doi: 10.1016/j.measurement.2022.111067.

[5] B. Zhang and S. Wang, "Analysis of the Susceptibility of Capacitive Touchscreens to External Electric Field Interference," in 2022 Asia-Pacific International Symposium on Electromagnetic Compatibility (APEMC), Sep. 2022, pp. 739–741. doi: 10.1109/APEMC53576.2022.9888500.

[6] G. Barrett and R. Omote, "Projected-Capacitive Touch Technology," Inf. Disp., vol. 26, no. 3, pp. 16–21, Mar. 2010, doi: 10.1002/j.2637-496X.2010.tb00229.x.

[7] L. E. Crooks, "Noise reduction techniques in electronic systems (2nd ed.), Henry W. Ott. Wiley-Interscience, New York. 1988," Magn. Reson. Med., vol. 10, no. 3, pp. 426–427, Jun. 1989, doi: 10.1002/mrm.1910100315.

[8] K. R. A. Britto, R. Dhanasekaran, R. Vimala, and B. Saranya, "Modeling of conducted EMI in flyback switching power converters," in 2011 INTERNATIONAL CONFERENCE ON RECENT ADVANCEMENTS IN ELECTRICAL, ELECTRONICS AND CONTROL ENGINEERING, Sivakasi, India: IEEE, Dec. 2011, pp. 377–383. doi: 10.1109/ICONRAEeCE.2011.6129801.

[9] S. Malik, M. Ahmad, S. Laxmeesha, T. Islam, and M. S. Baghini, "Impedance-to-Time Converter Circuit for Leaky Capacitive Sensors With Small Offset Capacitance," IEEE Sens. Lett., vol. 3, no. 7, pp. 1–4, Jul. 2019, doi: 10.1109/LSENS.2019.2919894.

[10] C. Riener et al., "Broadband Modeling and Simulation Strategy for Conducted Emissions of Power Electronic Systems Up to 400 MHz," Electronics, vol. 11, no. 24, Art. no. 24, Jan. 2022, doi: 10.3390/electronics11244217.

[11] Ling Jiang, Frank Wang, Keith Szolusha, and Kurk Mathews, "A practical method for separating common-mode and differential-mode emissions in conducted emissions testing." Analog Devices, Mar. 30, 2021. [Online]. Available: https://www.analog.com/en/analog-dialogue/articles/separating-common-mode-and-differential-mode-emissions-in-conducted-emissions-testing.html.

[12] H. Kim and B.-W. Min, "A study on EMI generation from a capacitive touch screen panel," in 2017 Asia-Pacific International Symposium on Electromagnetic Compatibility (APEMC), Jun. 2017, pp. 344–346. doi: 10.1109/APEMC.2017.7975501.

[13] M. I. Montrose, EMC and the Printed Circuit Board: Design, Theory, and Layout Made Simple. in IEEE Press Series on Electronics Technology. Wiley, 2004. [Online]. Available: https://books.google.cz/books?id=N-doGCRr2cAC.